\documentclass[twocolumn]{jpsj2} 
\usepackage{bm}

\title{Quantum Monte Carlo Study of the Quasi-One-Dimensional
Superconductivity} 

\author{Tohru \textsc{Aonuma}
,
Yuki \textsc{Fuseya} 
and 
Masao \textsc{Ogata}
}

\inst{
Department of Physics, University of Tokyo, Hongo, Bunkyo-ku,
Tokyo 113-0033
}

\abst{
We report on the result of quantum Monte Carlo simulation of
quasi-one-dimensional electron systems 
at $1/4$-filling,
considering organic superconductors such as TMTSF- and TMTTF-salts.
We focus on the effect of dimensionality (interchain coupling) on
superconducting fluctuation. 
First we consider Hubbard model which includes only on-site
repulsion $U$. 
We find that the increase of interchain coupling enhances
superconducting susceptibility, although it 
deforms the nested Fermi surface and 
suppresses the spin
susceptibility. 
Next we consider an extended Hubbard model which includes
nearest-neighbor repulsion $V$. 
In this case we consider the competition between different symmetries of
electron pairing,
$d_{x^2-y^2}$- and $d_{xy}$-wave symmetry, 
and find that 
the particle-particle interaction vertex for $d_{x^2-y^2}$-wave pairing
is no longer attractive 
even in the region where nesting
condition still holds. On the other hand, the interaction vertex for
$d_{xy}$-wave pairing persists to be attractive. 
The obtained results for the Hubbard model show qualitative agreement with recent
renormalization group study, and 
have the possibility
of accounting for the recent experimental result of (TMTTF)$_2$SbF$_6$, 
which exhibits a wider superconducting phase than the previous studies.
}

\kword{superconductivity, quasi-one-dimension, extended Hubbard model,
quantum Monte Carlo, organic conductors, TMTSF, TMTTF} 

\newcommand{\iu}{\mathrm{i}}
\newcommand{\e}{\mathrm{e}}
\newcommand{\expect}[1]{\langle #1 \rangle}
\usepackage{color}
\newcommand{\yf}[1]{\textcolor{red}{#1}\marginpar{\textcolor{red}{$\surd$}}}
\renewcommand{\yf}[1]{#1}

\begin{document}
\maketitle

\section{Introduction} 

Quasi-one-dimensional (Q1D) electron systems, such as organic
conductors TMTSF- and
TMTTF-salts \yf{(TM$_2$X)}, have been extensively
studied both theoretically and experimentally after the
discovery of superconductivity in (TMTSF)$_2$PF$_6$
\cite{JPhysLett.41.L95}. 
Various experiments have been carried out and the results 
are summarized as the so-called J\'{e}rome's phase diagram
\cite{Science.252.1509}. In 
this generic diagram, the ground state changes as spin-Peierls,
spin-density-wave (SDW) and superconductivity (SC) by applying pressure.
The SC phase is located next to the SDW phase; it suggests that the
source of the attractive interaction between electrons is due to 
antiferromagnetic spin fluctuation. The fact that the SC 
transition temperature, $T_{\rm c}$, decreases as applying pressure
(going away from SDW phase) supports this prediction. 
Many theoretical studies have been performed from this point of view
\yf{
\cite{JPSJ.58.1735,JPSJ.68.1481,JLowTempPhys.117.317,Emery,Duprat}
}
and
the obtained results give qualitative agreement with the experiments. 

A recent experimental result exhibits, however, a wider SC
phase under high-pressure \cite{JPSJ.77.023701}. This suggests that the
effect of pressure (or increase of interchain hopping) on SC phase is 
positive. From the theoretical point of view, the possibility of higher
$T_{\rm c}$ due to the increase of interchain hopping has been proposed
from the early days on the basis of one-dimensional (1D)
renormalization group theory \cite{J.Low.Temp.Phys.31.273}.  
Recently, 
modification of this approach makes it possible
to treat strong quantum fluctuations in Q1D system 
\cite{JPSJ.76.014709}, 
and concludes that the $d_{x^2-y^2}$-wave 
pair-field susceptibility shows clear enhancement against interchain
hopping 
\cite{Fuseya05,JPSJ.76.093701}.
It indicates that the decline of 
$T_{\rm c}$ along pressure may not be the case.

However, the previous studies are mainly based on the perturbative
methods, especially the ones which focus on the spin fluctuation.
Although it has been verified that the results of these approaches
qualitatively agree
with that of numerical methods in two-dimensional systems 
\cite{PhysRevB.43.8044}, 
it has not been confirmed yet whether 
these 
approximations are good 
in 1D or Q1D systems. 
The main obstacle is the strong quantum fluctuations in these systems,
i.e., 
spin fluctuation and 
the other types of
fluctuations are competitive and they interfere with each
other. 
One of the methods which can treat these fluctuations correctly is the
Q1D renormalization group theory mentioned above. 
However, it has its origins in the 1D
renormalization and is applicable only to weakly-coupled chains. 
Thus, in this paper, we adopt the quantum Monte Carlo (QMC) method 
in order to 
treat not only the spin fluctuation but also all types of
fluctuations at the same weight, which is rather difficult in 
perturbative approaches such as RPA or FLEX approximation. 
In addition, it can treat any strength of interchain coupling without
approximation in contrast to the Q1D renormalization group theory. 
We investigate the effect of interchain hopping, or dimensionality, 
on the SC fluctuation in the Hubbard 
model by using the auxiliary field QMC, \cite{PhysRevB.31.4403,
PhysRevB.40.506} having in mind 
the SC phase of TM$_2$X salts under pressure.

Furthermore, we study the effect of long range Coulomb interaction $V$,
\yf{
which has been recognized to be very important in organic conductors, especially the TMTTF-salts, in the context of the charge ordering (CO)\cite{Seo1,Seo2}. 
So far, the effect of $V$ on the SC has been investigated for the SC phase next to the CO phase, such as $\alpha$-ET$_2$I$_3$\cite{Kobayashi1,Kobayashi2,Kobayashi3}, or $\beta$-(DMeET)$_2$PF$_6$\cite{Yoshimi}.
As for TM$_2$X salts, on the other hand, it has not been studied since the SC phase is separated from the CO phase.
However, the high-energy peak ($\sim 30$meV) of the optical conductivity has been observed even in the metallic TMTSF$_2$X salts\cite{Science.281.1181}, indicating a precursor of CO.
The effect of $V$, therefore, is not negligible even in the metallic Q1D compounds;
it can give an sizable impact on the mechanism of SC.
In order to clarify this effect,
}
we also investigate the behavior of SC susceptibility for the Q1D extended
Hubbard model. 
%


This paper is organized as follows. 
In \S 2, we introduce the model
and method used in this paper. 
In \S 3, the
results of the Hubbard model 
are presented, which correspond to the situation
similar to the previous studies. They show qualitative agreement with
the renormalization group study \cite{JPSJ.76.093701}. In
\S 4, the result of the extended Hubbard model 
is presented. The effect of $V$ mainly appears
as the change of SC symmetry from $d_{x^2-y^2}$ to $d_{xy}$. 
The comparison with the experimental results is given in \S 5.

\section{Model and Method}
\begin{figure}[htbp]
 \begin{center}
  \includegraphics[width=3cm]{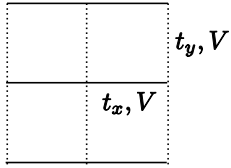}
 \end{center}
 \caption{Transfer integrals $t_x$, $t_y$ and nearest-neighbor repulsion 
 $V$. }
 \label{latticeV}
\end{figure}

We consider the two-dimensional Hubbard and extended Hubbard model on an
anisotropic square lattice, whose Hamiltonian is given by 
\begin{align}
 H = & \ 
 \sum_{\expect{ij}\sigma} \left(
 -t_{ij} c_{i\sigma}^\dag c_{j\sigma} + \text{h.c.}
 \right)
 + U \sum_{i} n_{i\uparrow} n_{i\downarrow} \notag \\
 & + V \sum_{\expect{ij}\sigma\sigma'} n_{i\sigma}n_{j\sigma'}
 - \mu \sum_{i\sigma} n_{i\sigma},
 \label{EHM}
\end{align}
where 
$c_{i\sigma}^\dag$ ($c_{i\sigma}$) is
the creation (annihilation) operator of an electron with spin $\sigma$
($\sigma=\uparrow,\downarrow$) on site $i$ and
$n_{i\sigma}=c_{i\sigma}^\dag c_{i\sigma}$. 
Here 
$t_{ij}$ has two different values, $t_x$ (intrachain hopping) and 
$t_y$ (interchain hopping). 
The nearest-neighbor Coulomb interaction $V$ is
introduced as depicted in Fig. \ref{latticeV}. 
We set the intrachain hopping $t_x=1$, 
interaction strength $U=4$ and 
$k_{\rm B}=\hbar=1$ throughout this paper. 
The number of electrons are set to be $\expect{n}=1/4$ by tuning the 
chemical potential $\mu$.

In order to investigate the properties of this model, we employ the
auxiliary field quantum Monte Carlo method \cite{PhysRevB.31.4403,
PhysRevB.40.506}. 
The shape of the cluster is set to be $16 \times 8$ for $V=0$ or 
$16 \times 4$ for $V\neq0$. It
means that there 
are 16 sites along chains ($x$-direction) and 8 (or 4) chains are coupled with
each other in the $y$-direction. 
In the presence of $V$, 
we restrict the system 
size to $16\times4$ and the temperature range to $\beta (=1/T) \le 4$,  
because of
heavy computational cost of Monte Carlo 
sampling \cite{PhysRevB.39.9397}. 
The imaginary-time step is set to be $\Delta\tau=1/8$, and 
64000 
measurements separated by two sweeps are performed 
after 1000 warm-up. 

\begin{figure}[htbp]
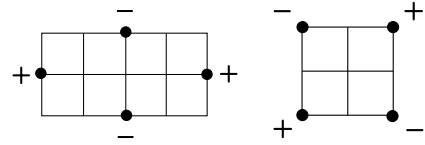

 \begin{center}
  \includegraphics[width=3cm]{fig02a.epsi} \quad 
  \includegraphics[width=2cm]{fig02b.epsi} 
 \end{center}
 \caption{The internal pair coordinate $\delta$ and the factor
 $f_d(\delta)$ used in defining $\Delta_{i}$ for $d_{x^2-y^2}$-wave
 pairing (left panel) and $d_{xy}$-wave pairing (right panel). }
 \label{fig.000}
\end{figure}

We calculate spin, charge and pair-field susceptibilities 
\begin{align}
 \chi_{\rm s} (\bm{q})
 = & \ 
 \frac{1}{N}
 \int_0^\beta {\rm d} \tau
 \sum_{ij}
 \e^{\iu\bm{q}\cdot(\bm{r}_i-\bm{r}_j)}
 \expect{m_i^z(\tau) m_j^z(0)}, \notag \\
 \chi_{\rm c} (\bm{q})
 = & \
 \frac{1}{N}
 \int_0^\beta {\rm d} \tau
 \sum_{ij}
 \e^{\iu\bm{q}\cdot(\bm{r}_i-\bm{r}_j)}
 \expect{\tilde{n}_i(\tau) \tilde{n}_j(0)}, \notag \\
 \chi_{d{\rm SC}}
 = & \
 \frac{1}{4N}
 \int_0^\beta {\rm d} \tau
 \sum_{ij}
 \expect{\Delta_{i}(\tau) \Delta_{j}^\dag(0)} \label{full}
\end{align}
with
\begin{align*}
 m_i^z(\tau)
 = & \ 
 c_{i\uparrow}^\dag (\tau)
 c_{i\uparrow} (\tau)
 -
 c_{i\downarrow}^\dag (\tau)
 c_{i\downarrow} (\tau), \\
 \tilde{n}_i(\tau)
 = & \ 
 c_{i\uparrow}^\dag (\tau)
 c_{i\uparrow} (\tau)
 +
 c_{i\downarrow}^\dag (\tau)
 c_{i\downarrow} (\tau)
 - 
 \expect{n}, \\
 \Delta_{i} (\tau)
 = & \ 
 \sum_{\delta} 
 f_{d}(\delta) 
 c_{i\uparrow}(\tau) 
 c_{i+\delta\downarrow} (\tau).
\end{align*}
Here $f_{d}(\delta)$ is the factor $+1$ or $-1$ corresponding to the
pairing symmetries as shown in Fig.
\ref{fig.000}. 

In addition to eq. (\ref{full}), we also calculate the uncorrelated
pair-field susceptibility\cite{JPSJ.74.1390, PhysRevB.72.134513}
\begin{align}
 \bar{\chi}_{d{\rm SC}}
 = & \
 \frac{1}{4N}
 \int_0^\beta {\rm d} \tau
 \sum_{i\delta j\delta'}
 f_d(\delta)
 f_d(\delta') \notag \\
 & \times
 \underbrace{
 \expect{c_{i\uparrow}(\tau) c_{j\uparrow}^\dag(0)}
 }_{ -G_{i,j,\uparrow}(\tau) }
 \underbrace{
 \expect{c_{i+\delta\downarrow}(\tau) c_{j+\delta'\downarrow}^\dag(0)}
 }_{ -G_{i+\delta,j+\delta',\downarrow}(\tau) }.
 \label{wo}
\end{align}
The difference between eqs. (\ref{full}) and (\ref{wo}) is whether the
particle-particle vertex $\Gamma$ is included or not (see Fig.
\ref{vertex}). 
\begin{figure}[htbp]
 \begin{center}
  \includegraphics[width=6cm]{fig03.epsi} 
 \end{center}
 \caption{Full susceptibility $\chi_{\rm SC}$ and uncorrelated susceptibility
 $\bar{\chi}_{\rm SC}$. Thick lines denote the dressed Green's
 functions. $\Gamma$ is the particle-particle interaction vertex.}
 \label{vertex}
\end{figure}
If $\chi_{\rm SC}$ is larger than $\bar{\chi}_{\rm SC}$, it means that 
$\Gamma$ 
is attractive \cite{JPSJ.74.1390, PhysRevB.72.134513, PhysRevB.39.839}.

We also calculate the density
of states $N(\omega)$, which is defined as 
\begin{equation}
 \sum_{i} G_{ii} (\tau)
 =
 - \int_{-\infty}^{\infty}
 {\rm d} \omega
 \frac{\e^{-\omega\tau}}{1+\e^{-\beta\omega}}
 N(\omega),
 \label{MaxEnt}
\end{equation}
where $G_{ij}(\tau)$ is the imaginary-time Green's function
\[
 G_{ij} (\tau)
 =
-
 \sum_{\sigma}
 \expect{c_{i\sigma}(\tau) c_{j\sigma}^\dag(0)}. 
\]
This expression will be readily understood from the 
definition of $N(\omega)$ 
\begin{gather*}
 N(\omega) 
 =
 \frac{1}{N}
 \sum_{\bm{k}} A(\bm{k}, \omega)
\end{gather*}
with $A(\bm{k},\omega)$ being the one-particle excitation spectrum
\[
 G(\bm{k}, \tau) 
 =
 -
 \int_{-\infty}^{\infty}
 {\rm d} \omega
 \frac{\e^{-\omega\tau}}{1+\e^{-\beta\omega}}
 A(\bm{k},\omega).
\]
The inverse transformation from $G_{ii}(\tau)$ to $N(\omega)$ in
eq. (\ref{MaxEnt}) is carried out by the maximum entropy method combined
with QMC \cite{PhysRep.269.133}.

\section{Results of the Hubbard model}

\begin{figure}[htbp]
 \begin{center}
  \includegraphics[width=8cm]{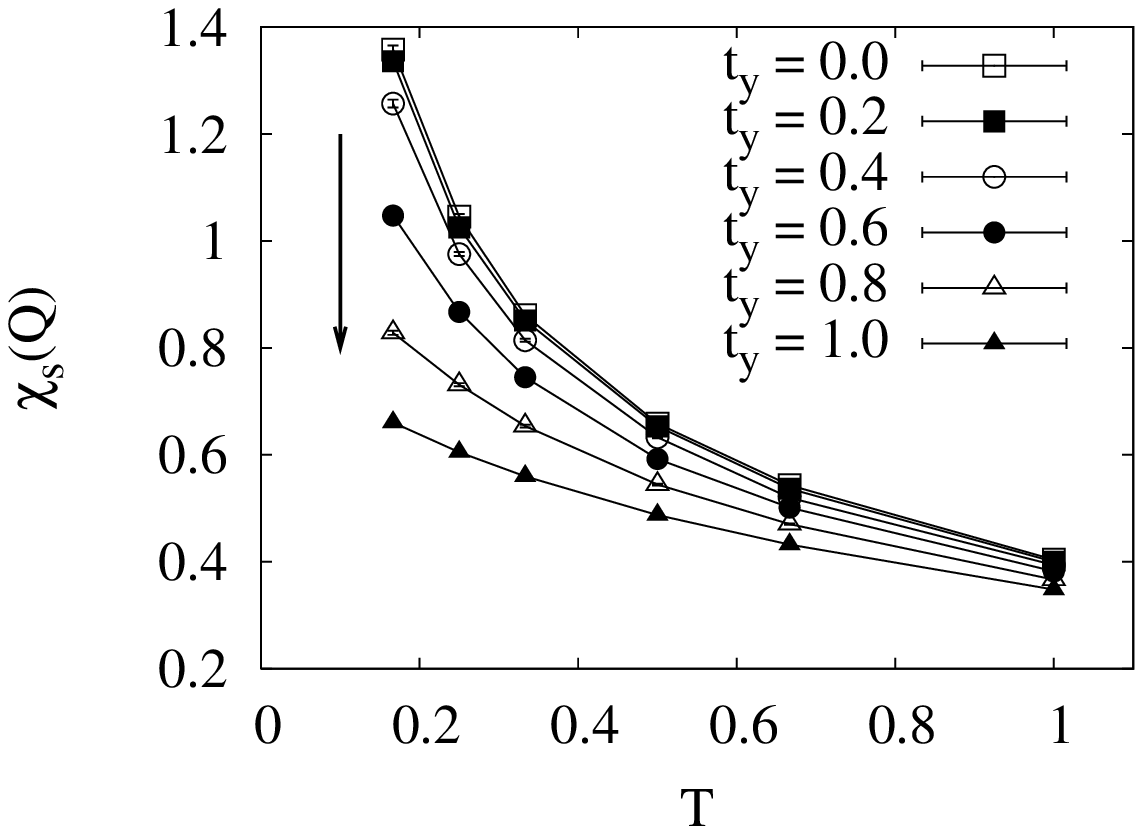} \\ 
  \includegraphics[width=8cm]{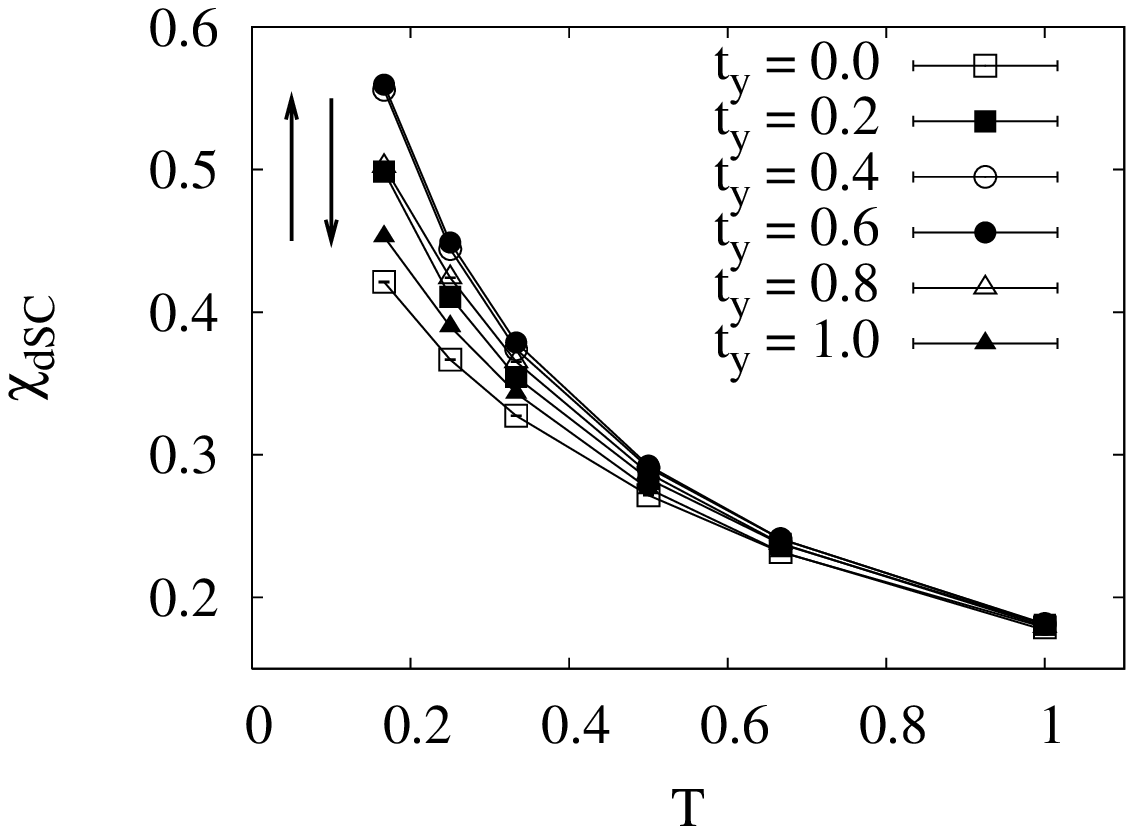} 
 \end{center}
 \caption{Temperature dependences of spin susceptibility 
 $\chi_{\rm s}(\bm{Q})$ (upper panel) and $d_{x^2-y^2}$-wave pair-field
 susceptibility $\chi_{d{\rm SC}}$ (lower panel) for $U=4$ and different values of
 $t_y$. The change of these susceptibilities when $t_y$ is increased are
 indicated by arrows. Where not shown, error bars are smaller than the
 symbols.}
 \label{Jan 25 14:35:27 2008}
\end{figure}

\begin{figure}[htbp]
 \begin{center}
  \includegraphics[width=5.5cm]{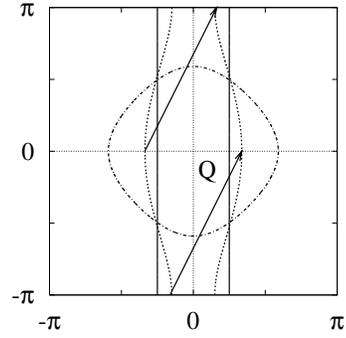} 
 \end{center}
 \caption{Fermi surfaces with $t_y=0.0$ (solid), $0.2$ (dotted) and
 $1.0$ (dash-dotted) at $1/4$-filling. The good nesting condition holds at
 about $0.0 \le t_y \le 0.3$. $\bm{Q}=(\pi/2,\pi)$ is a nesting vector.} 
 \label{FS}
\end{figure}

First we show the results of the Hubbard model, whose Hamiltonian is
given by eq. (\ref{EHM}) with $V=0$.
Figure \ref{Jan 25 14:35:27 2008} is 
the spin susceptibility $\chi_{\rm s}(\bm{Q})$ and SC susceptibility
$\chi_{d{\rm SC}}$ against temperature 
$T$ for different values of $t_y$. 
As for $\chi_{\rm s}(\bm{Q})$, it is strongly enhanced at low
temperature near 1D, but the enhancement becomes moderate 
as $t_y$ increases. 
This is due to the breaking of the nesting condition, 
as depicted in Fig. \ref{FS}.
On the other hand, $\chi_{d {\rm SC}}$ shows different behaviors. 
$\chi_{d{\rm SC}}$ is {\it enhanced} by $t_y$ in small $t_y$ region,
while
$\chi_{\rm s}(\bm{Q})$ 
decreases. 
$\chi_{d{\rm SC}}$ becomes largest at the intermediate values of
$t_y$, such as $t_y=0.4$ or $0.6$. 

\begin{figure}[htbp]
 \begin{center}
  \includegraphics[width=8cm]{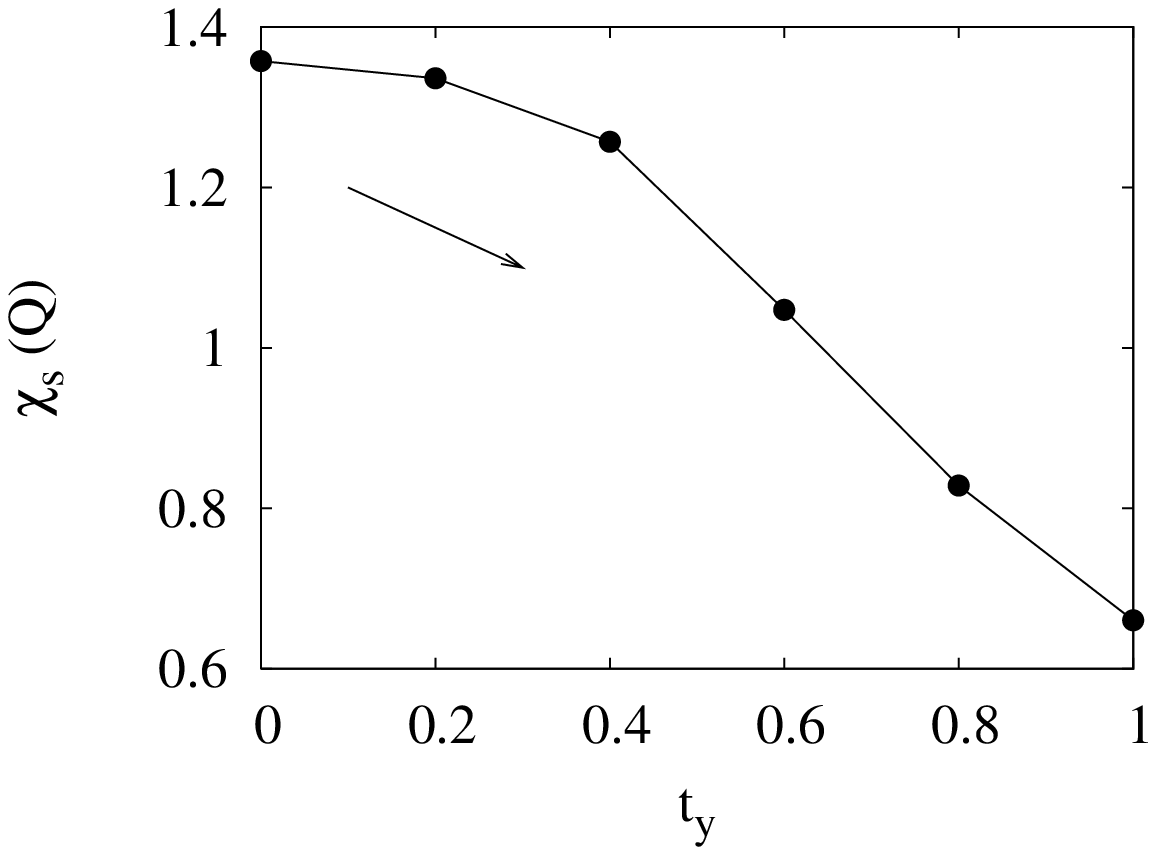} \\ 
  \includegraphics[width=8cm]{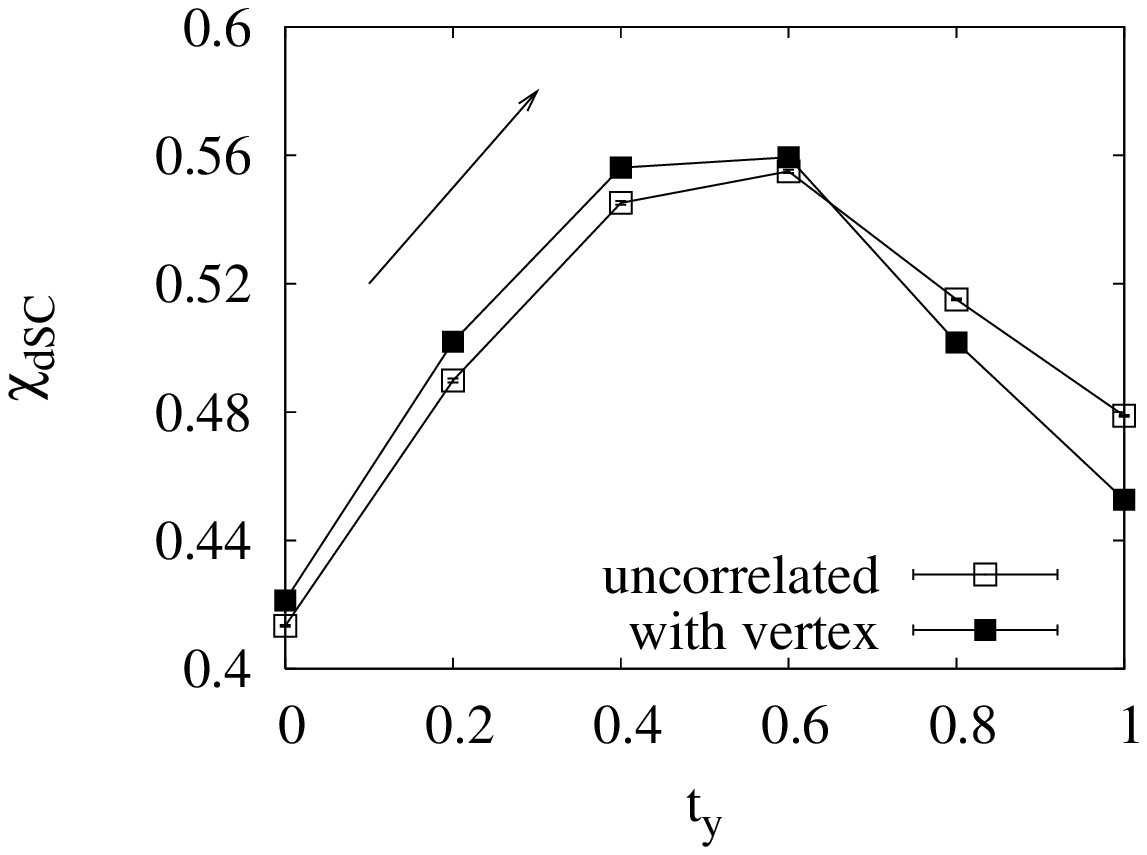} 
 \end{center}
 \caption{$t_y$-dependences of spin susceptibility 
 $\chi_{\rm s}(\bm{Q})$ (upper panel) and $d_{x^2-y^2}$-wave pair-field
 susceptibility $\chi_{d{\rm SC}}$ (lower panel, solid squares) at
 $T=0.167$ for $U=4$. The open squares denote the uncorrelated pair-field
 susceptibility $\bar{\chi}_{d{\rm SC}}$. 
 } 
 \label{Jan 25 14:50:48 2008}
\end{figure}

The $t_y$-dependences of $\chi_{s}(\bm{Q})$ and $\chi_{d{\rm SC}}$ are
summarized in Fig. 
\ref{Jan 25 14:50:48 2008} when the temperature is fixed at $T=0.167$. 
The non-monotonic behavior of $\chi_{d{\rm SC}}$ can
be interpreted as follows: 
the enhancement comes from the pair-hopping of electrons, and the
suppression is due to the suppression of spin fluctuation because of the
deformation of the nested Fermi surface. 
In fact, the peak position of $\chi_{\rm s}(\bm{q})$ locates at around
$\bm{q}=\bm{Q}$ for $0.0 \le t_y \le 0.6$, while it does not 
at $t_y=0.8$ and $1.0$ (not shown). 

In Fig. \ref{Jan 25 14:50:48 2008}, the uncorrelated susceptibility,
$\bar{\chi}_{d{\rm SC}}$, is also shown in comparison with the full
susceptibility, $\chi_{d{\rm SC}}$. We find that the vertex part is
attractive for $0.0 \le t_y \le 0.6$, while it is repulsive for
$t_y=0.8$ and $1.0$. This is consistent with the above picture 
that 
$\chi_{d{\rm SC}}$ is most enhanced at the region where the nesting
property and the interchain pair-hopping of electrons are well
balanced. 

\begin{figure}[htbp]
 \begin{center}
  \includegraphics[width=2.5cm]{fig07a.epsi} 
  \includegraphics[width=5.5cm]{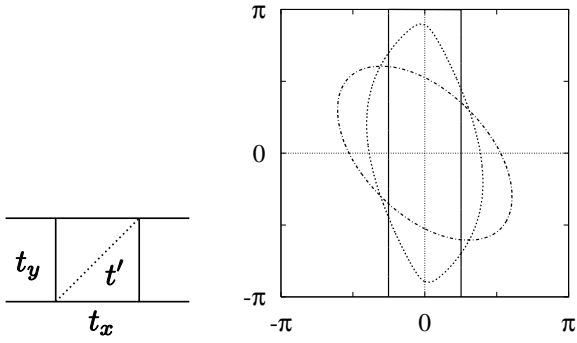} 
 \end{center}
 \caption{Left panel: the interchain hopping $t'$, which induces
 geometrical frustration.
 Right panel: Fermi surfaces with frustration $t'$ when $t_y=0.0$
 (solid), $t_y=0.2$ (dotted) and $1.0$ (dash-dotted) at
 $1/4$-filling. The value of $t'$ is chosen as $t'=t_y/2$. It should be
 noted that the nesting vector $\bm{Q}$ no longer exists.} 
 \label{frustration}
\end{figure}

In order to confirm the effect of nesting property, it is useful to
introduce the frustrated hopping, $t'$, as shown in
Fig. \ref{frustration}. 
This kind of hopping is 
another key to control the strength of spin fluctuation and interchain
coupling. 
In fact, 
this type of transfer integral has been confirmed to exist 
in many TM$_2$X-salts \cite{J.Phys.C.19.3805}. 
It deforms the Fermi surface and worsen the nesting condition (compare
Fig. \ref{FS} and \ref{frustration}).

\begin{figure}[htbp]
 \begin{center}
  \includegraphics[width=8cm]{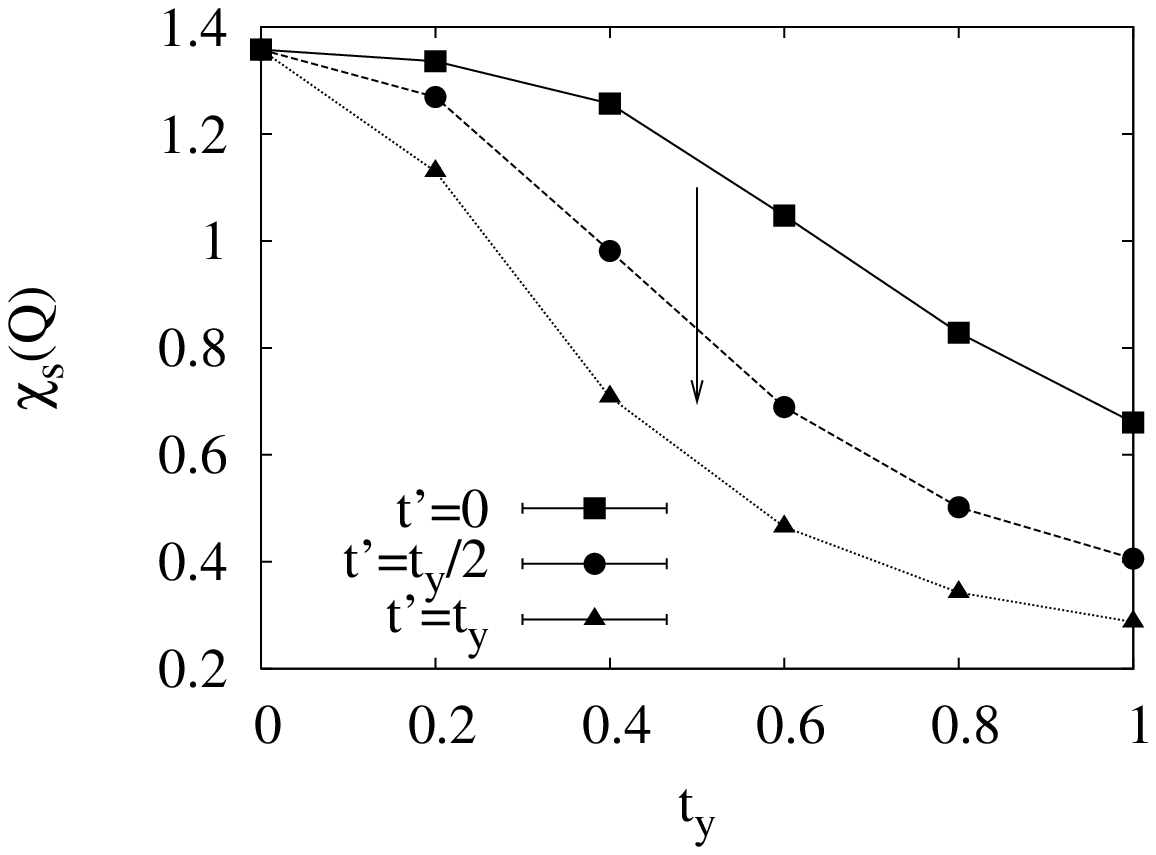} \\ 
  \includegraphics[width=8cm]{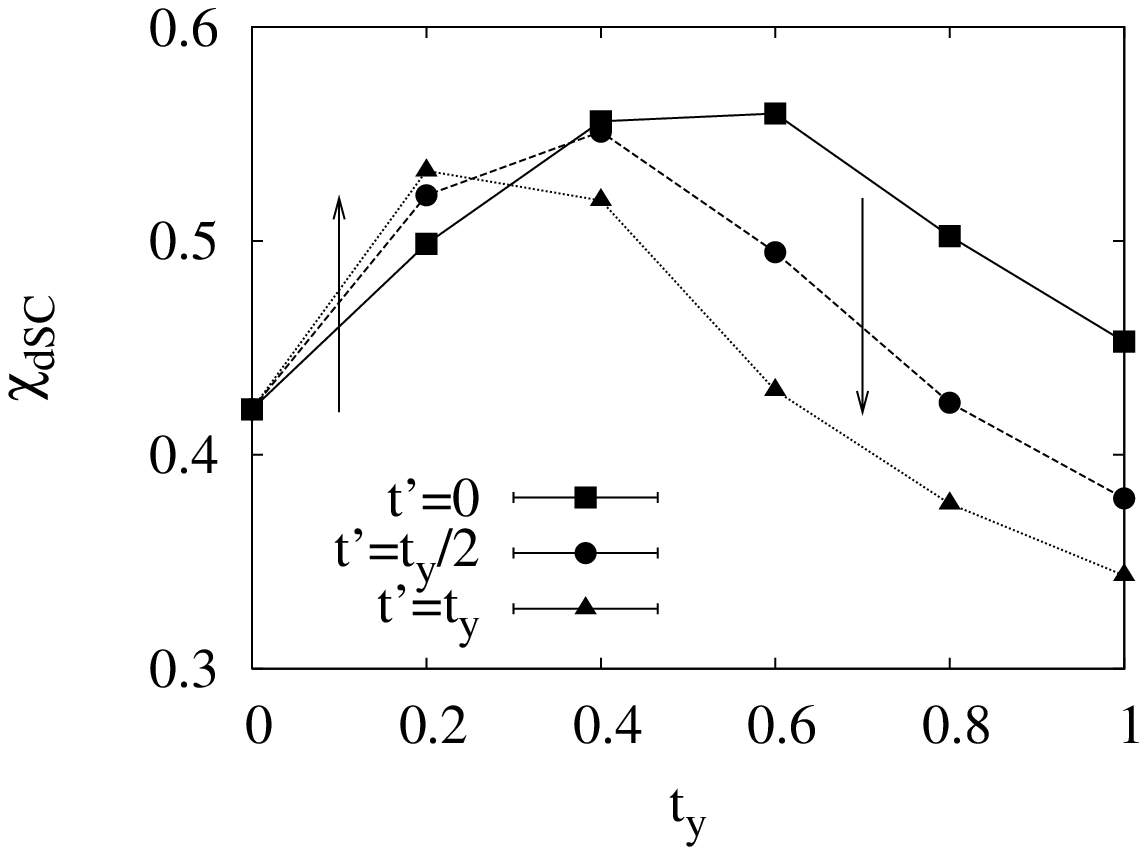} 
 \end{center}
 \caption{$t_y$-dependences of spin susceptibility 
 $\chi_{\rm s}(\bm{Q})$ (upper panel) and $d_{x^2-y^2}$-wave pair-field
 susceptibility $\chi_{d{\rm SC}}$ (lower panel) at
 $T=0.167$ for $U=4$. Squares, circles and triangles correspond to the case
 $t'=0$, $t_y/2$ and $t_y$. The change of these susceptibilities when
 $t'$ is strengthened are indicated by arrows.}
 \label{chi_frustrated}
\end{figure}

Figure \ref{chi_frustrated} shows the $t_y$-dependence of spin and SC
susceptibilities 
at $T=0.167$ for different strengths of $t'$. 
Here we 
choose the values of $t'$ as $t'=0$, $t_y/2$ and $t_y$. It is seen that
$\chi_{d {\rm SC}}$ is slightly enhanced at small $t_y$ region and
suppressed at large $t_y$ region, 
while $\chi_{\rm s}(\bm{Q})$ is monotonically suppressed by $t'$. 
The 
enhancement of $\chi_{d{\rm SC}}$ will be again due to 
the increase of
interchain coupling. Indeed, 
the interchain hopping $t'$ 
couples the chains more tightly, 
while 
it worsen the nesting condition. 
It will increase the pair-hopping of electrons and, as a result, 
$\chi_{d{\rm SC}}$ is slightly enhanced by $t'$.
In addition, the region where 
the interaction vertex is attractive 
($\chi_{d {\rm SC}}-\bar{\chi}_{d {\rm SC}}>0$)
shifts to the smaller $t_y$ region (not shown), which simply supports this
picture. 

\begin{figure}[htbp]
 \begin{center}
  \includegraphics[width=7cm]{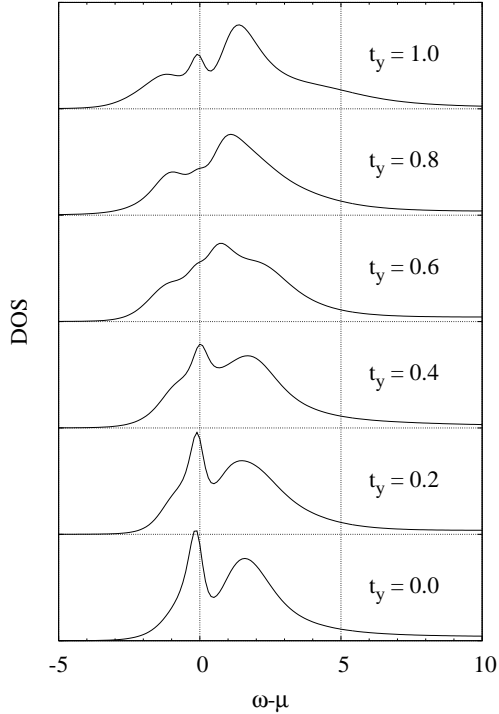} 
 \end{center}
 \caption{Density of States for $U=4$ and different values of $t_y$ at 
 $T=0.167$. The frustration $t'$ is not included.}
 \label{dos}
\end{figure}

However, one might point out that the maximum of $\chi_{d{\rm SC}}$ is
due to  
the van Hove singularity which
exists at $t_y \sim 0.35$ in $1/4$-filled non-interacting system, 
i.e., the effect of electron-electron interaction does not matter. 
In order to clarify 
this point, 
we calculate the density
of states (DOS) $N(\omega)$ for interacting system.
Figure \ref{dos} shows
$N(\omega)$ at $T=0.167$ for different values of
$t_y$ (here we consider the case $t'=0$). 
Although the suppression of $N(\omega=\mu)$ at large $t_y$ region is
recognized, the enhancement at small $t_y$ region seems not to exist.
This behavior 
does not follow the curve 
of $\chi_{d {\rm SC}}$ against $t_y$. 
Thus we conclude that the effect of van
Hove singularity is irrelevant 
to understand the non-monotonic
behavior of $\chi_{d{\rm SC}}$, 
as long as
the temperature range we calculated. 

\section{Results of the extended Hubbard model}

In the previous section, we investigate 
the case with $V=0$, where 
$\chi_{d{\rm SC}}$ is affected by the pair-hopping
of electrons, in addition to the strength of spin fluctuation. 
On the other hand, the pairing mechanism itself is well understood by
the spin-fluctuation theory. 
In this section, we consider whether this pairing mechanism is valid
or not under the presence of charge fluctuation induced by the
nearest-neighbor repulsion $V$.

We consider the extended Hubbard model, 
whose Hamiltonian is given by eq. (\ref{EHM}). 
We impose the limitation
on not only the system size and temperature but also $V$ as $V\le0.5$ 
in order to avoid round-off errors and severe sign problem. 
(We adopt the parameters which assure that the average sign
$\expect{S}$ becomes greater than $0.1$.) 

\begin{figure}[htbp]
 \begin{center}
  \includegraphics[width=4cm]{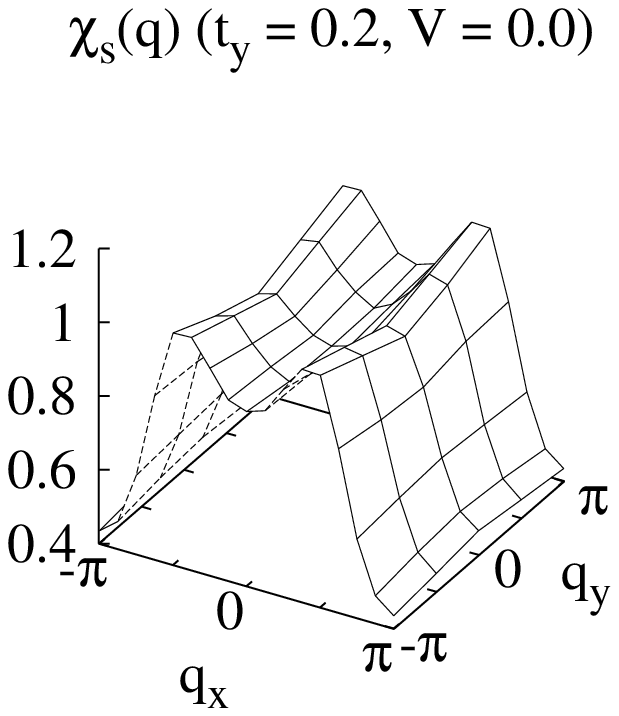} 
  \includegraphics[width=4cm]{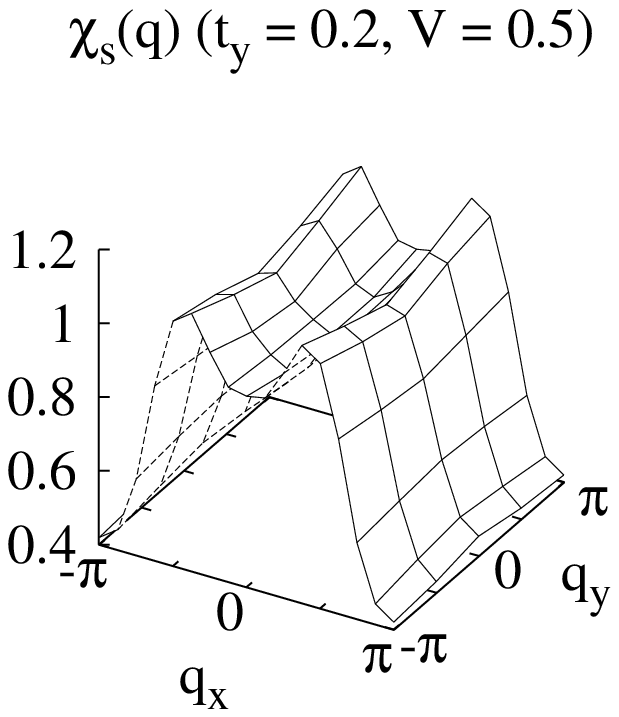} \\ 
  \includegraphics[width=4cm]{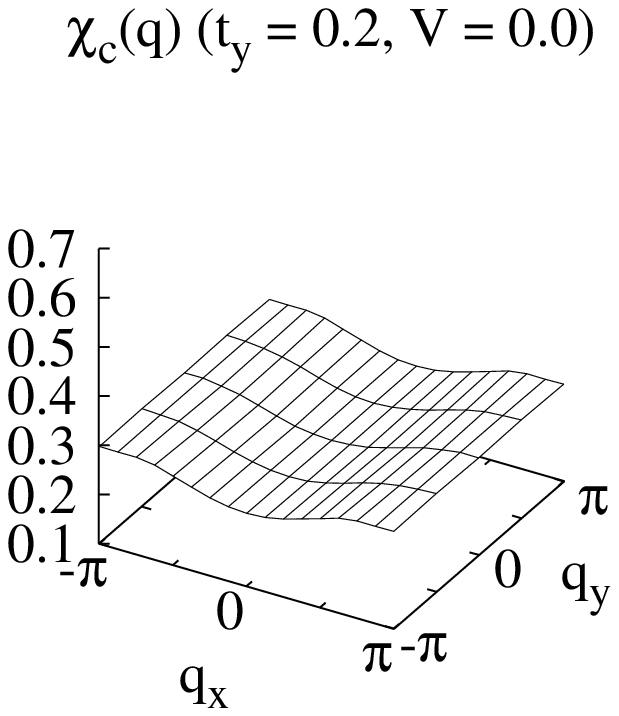} 
  \includegraphics[width=4cm]{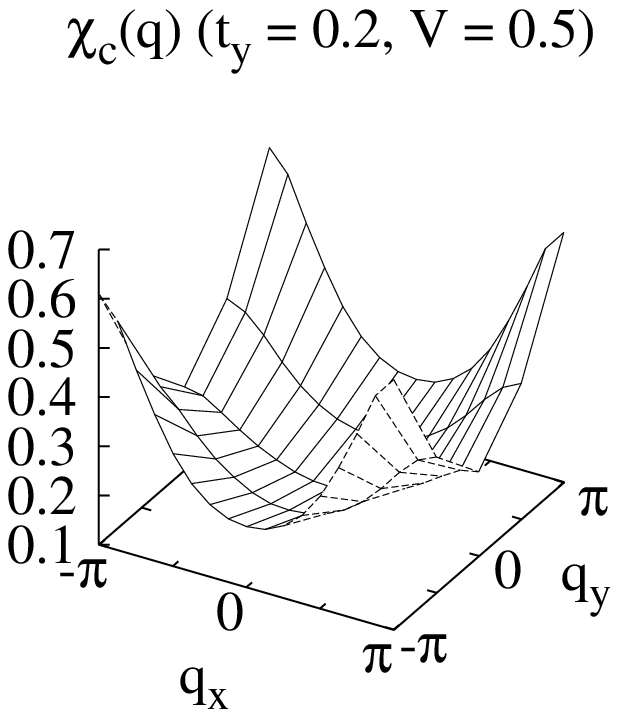} 
 \end{center}
 \caption{Spin susceptibility $\chi_{\rm s}(\bm{q})$ (upper panels) and
 charge susceptibility $\chi_{\rm c}(\bm{q})$ (lower panels) when $T=0.25$ and
 $t_y=0.2$ in the 1st Brillouin zone. The interaction strengths are $U=4$, $V=0$ (left
 panels) and $U=4$, $V=0.5$ (right panels). The structure of 
 $\chi_{\rm s}(\bm{q})$ reflects the 
 shape of the nested Fermi surface with a nesting vector $\bm{Q}=(\pi/2,\pi)$ and seems
 not to be affected by $V$. On the other hand, 
 the structure of $\chi_{\rm c}(\bm{q})$ is largely
 affected by $V$ and its peak position reflects the geometrical
 structure of $V$ (or its Fourier transform 
 $V(\bm{q}) \propto \cos(q_x) + \cos(q_y)$). 
 Note that
 each point includes errors with a few percent.}
 \label{chi_spin_charge}
\end{figure}

First we show the spin and charge susceptibilities with and without
$V$. 
The upper two panels of Fig. \ref{chi_spin_charge} represent the spin
susceptibility, $\chi_{\rm s}(\bm{q})$, for 
$V=0$ and $0.5$ when $T=0.25$ and $t_y=0.2$, where the temperature is
relatively low and the nesting condition still holds.  
It is seen that the structure of 
$\chi_{\rm s}(\bm{q})$ is hardly affected by $V$. It comes from the fact
that $\chi_{\rm s}(\bm{q})$ is mainly affected by the shape of the Fermi
surface, which is considered not to be drastically deformed by $V$. 
The lower two panels of Fig. \ref{chi_spin_charge} show 
the charge susceptibility, $\chi_{\rm c}(\bm{q})$.
A sharp peak at
$\bm{q}=(\pi,\pi)$ is formed when $V$ is finite, which is considered to
be the reflection of the lattice structure. (Note that we consider the
$1/4$-filled system, which favors $\bm{q}=(\pi,\pi)$ charge
disproportionation under finite $V$.) 
This result suggests that the charge fluctuation induced by $V$ is more affected
by the geometrical configuration of $V$ rather than the shape of the
Fermi surface \cite{footnote01}. 
It is a clear contrast between $\chi_{\rm s}$ and
$\chi_{\rm c}$: the nesting property 
(momentum space) and the
lattice structure (real space). 

\begin{figure}[htbp]
 \begin{center}
  \includegraphics[width=8cm]{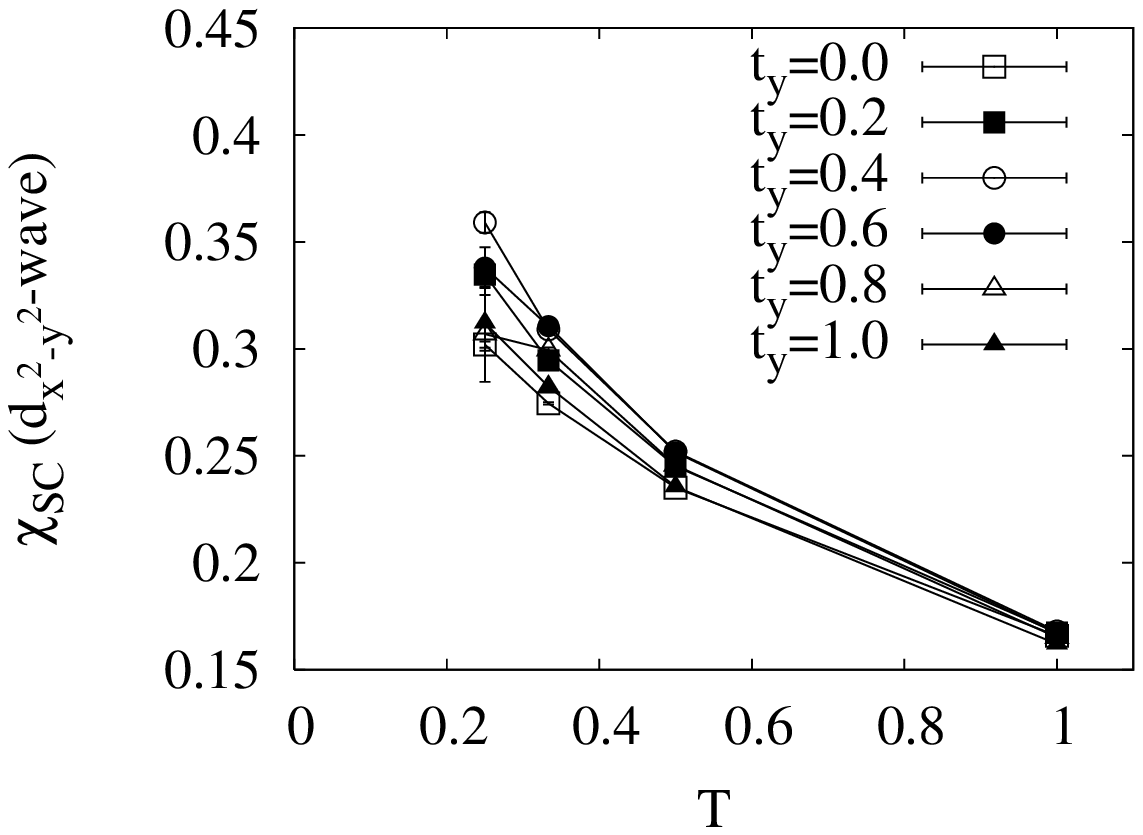} 
  \includegraphics[width=8cm]{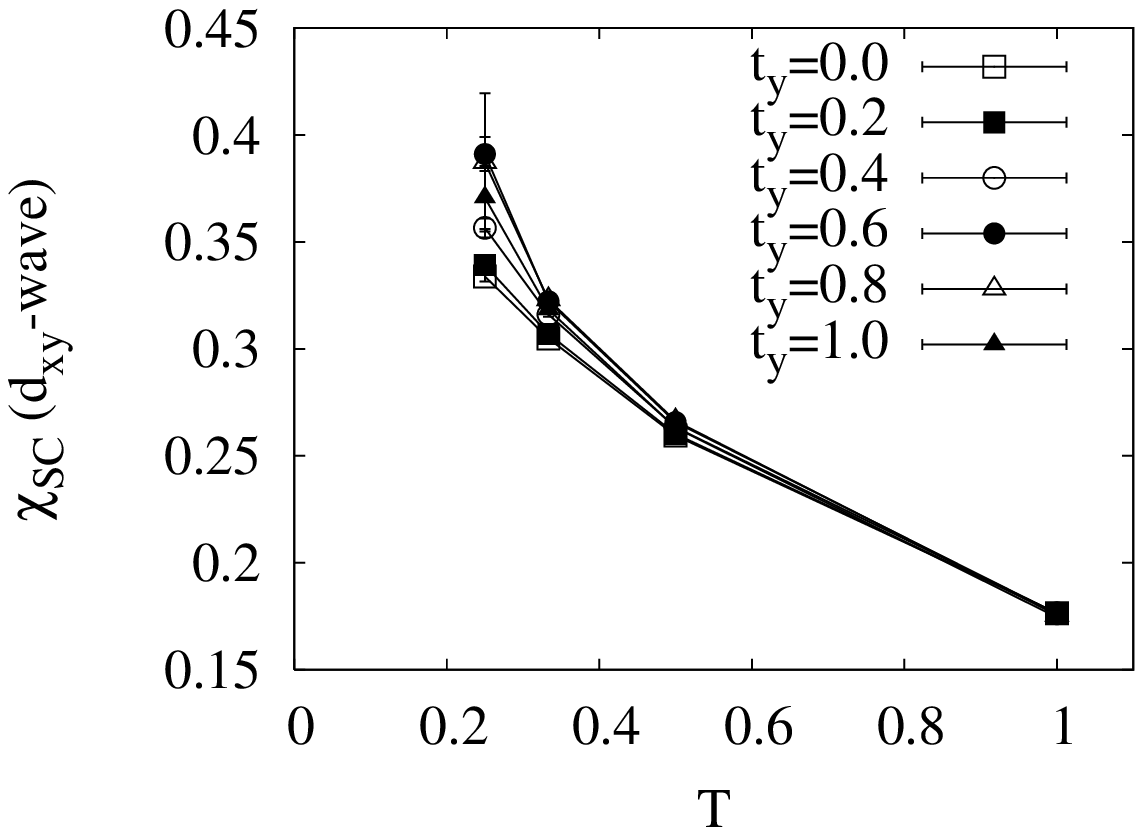} 
 \end{center}
 \caption{Temperature dependences of SC susceptibility for
 $d_{x^2-y^2}$-wave (upper panel) and $d_{xy}$-wave (lower panel)
 pairings for $\expect{n}=0.5$, $U=4$ and $V=0.5$.}
 \label{Feb 14 19:03:41 2008}
\end{figure}

\begin{figure}[htbp]
 \begin{center}
  \includegraphics[width=8cm]{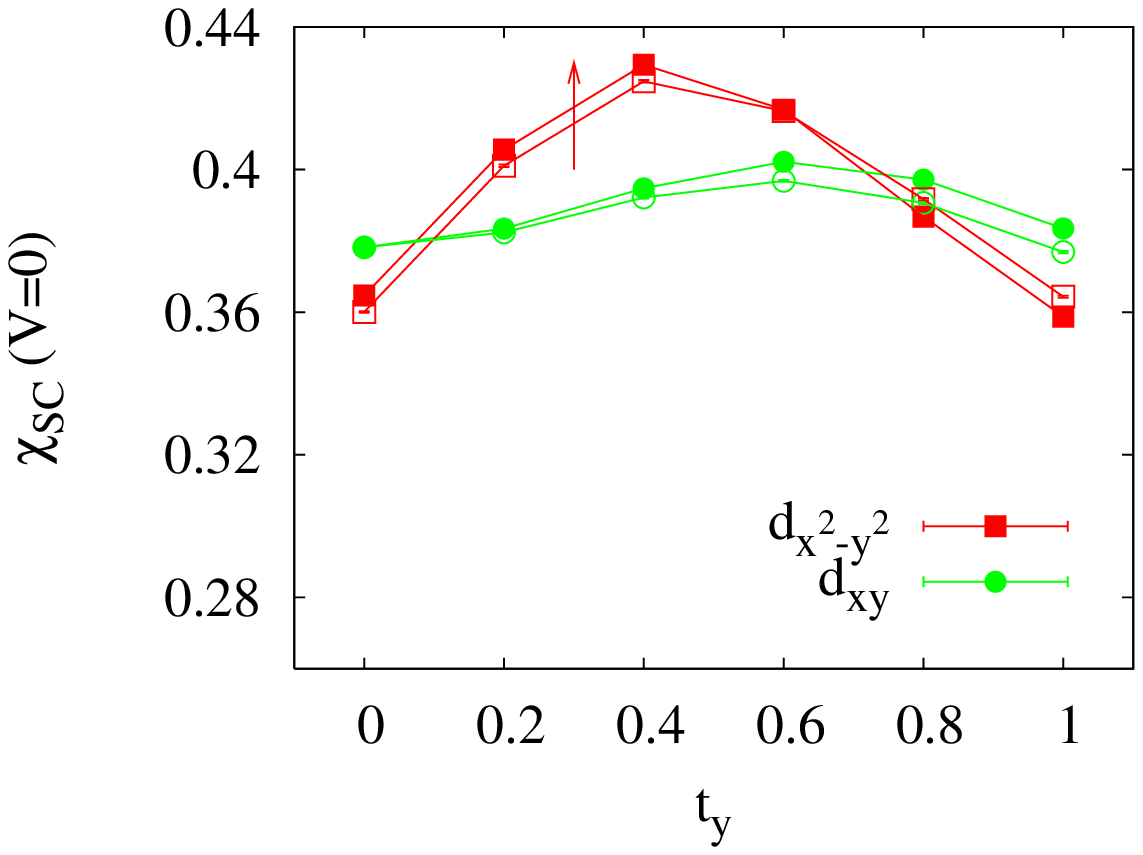} 
  \includegraphics[width=8cm]{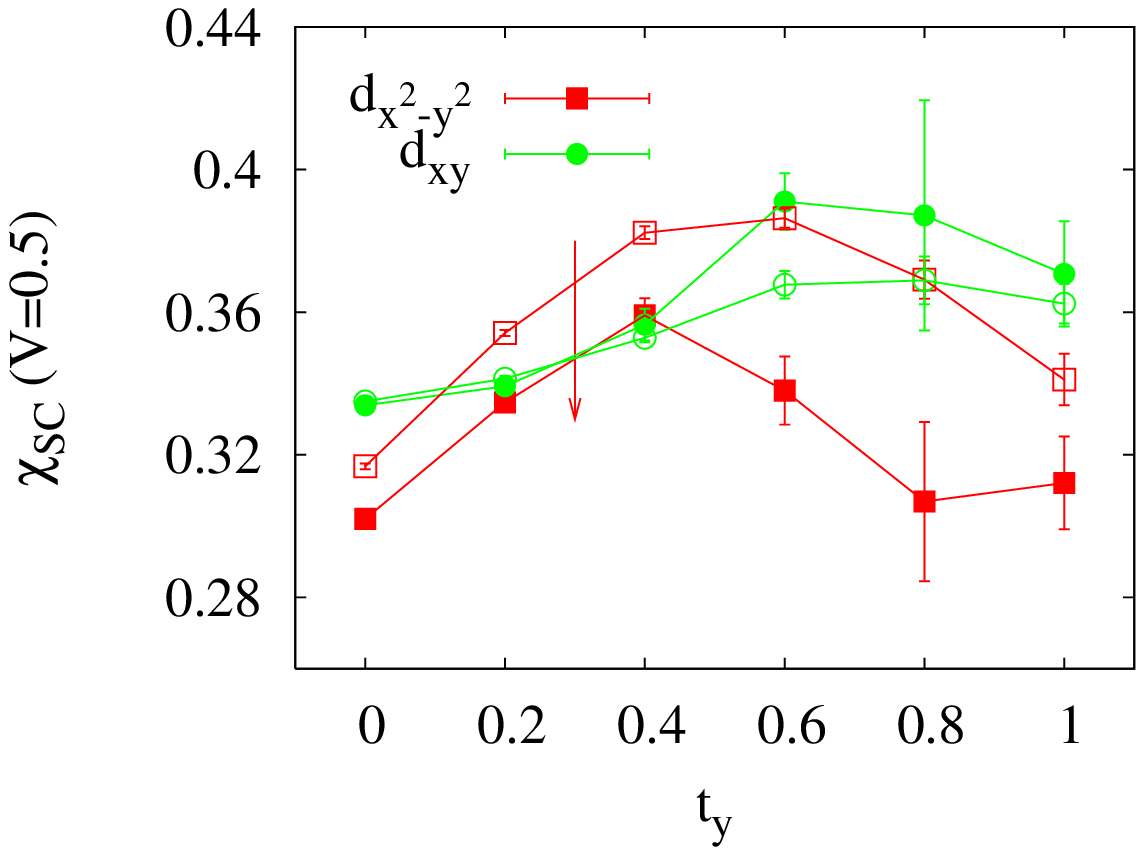} 
 \end{center}
 \caption{(Color online) $t_y$-dependence of SC susceptibility at $T=0.25$ for $U=4$, $V=0.0$
 (upper panel) and $U=4$, $V=0.5$ (lower panel). 
 The solid and open symbols denote the full and uncorrelated
 susceptibility. The contribution of particle-particle interaction
 vertex for $d_{x^2-y^2}$-wave pairing is indicated by arrows:
 attractive (upper panel) and repulsive (lower panel).
}
 \label{Feb 14 19:30:11 2008}
\end{figure}

It is an interesting subject to investigate how 
$V$ affects SC susceptibility. To understand the effect of charge
fluctuation, we  
consider not only the 
$d_{x^2-y^2}$-wave pairing but also $d_{xy}$-wave pairing, whose gap
function 
$\Delta_i$ is defined in 
Fig. \ref{fig.000}. 
Figure \ref{Feb 14 19:03:41 2008} shows the temperature dependence of
the SC susceptibilities for these two pairings. 
They are qualitatively similar, but
it seems that $d_{xy}$-wave pairing 
is slightly larger than the 
$d_{x^2-y^2}$-wave pairing. 
This advantage of $d_{xy}$-wave pairing over $d_{x^2-y^2}$-wave pairing
will be recognized more 
clearly in Fig. \ref{Feb 14 19:30:11 2008}, where the $t_y$-dependences
of these susceptibilities at $T=0.25$ are shown. 
(The results without $V$ are also displayed
for comparison.)
The $d_{x^2-y^2}$-wave pairing is drastically suppressed by
$V$, although the suppression of $d_{xy}$-wave pairing is not so large. 
Furthermore, it should be noted that, for $V=0.5$, the interaction vertex for
$d_{x^2-y^2}$-wave pairing is repulsive for {\it any} value $t_y$, even
for $t_y=0.0$ or $0.2$ where strong spin fluctuation still exists. This 
suggests that the spin and charge fluctuations cancel each other and
the vertex part for $d_{x^2-y^2}$-wave pairing is no longer
attractive. On the other hand, the interaction vertex for the
$d_{xy}$-wave pairing is still attractive; it is even enhanced at the region
of large $t_y$. 

\begin{figure}[htbp]
 \begin{center}
  \includegraphics[width=7cm]{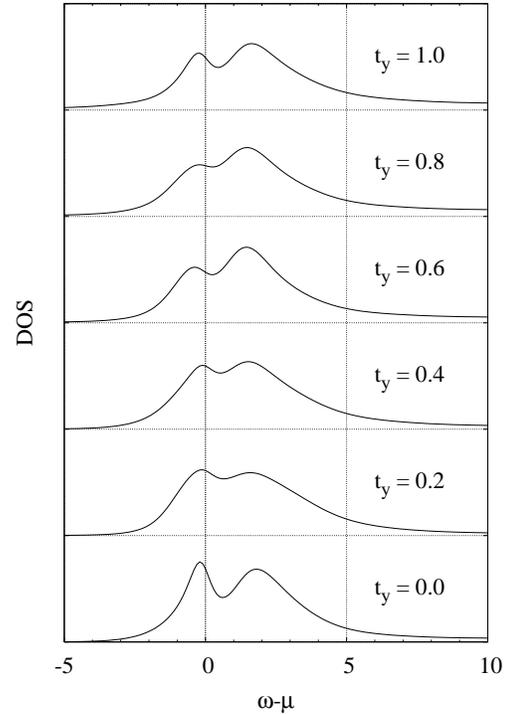} 
 \end{center}
 \caption{Density of states at $T=0.25$ for $U=4$ and $V=0.5$.}
 \label{dosv}
\end{figure}

Finally, 
we estimate the effects of $V$ on the dynamical property of this model. 
Figure \ref{dosv} shows the DOS, $N(\omega)$, for different
values of 
$t_y$ at $T=0.25$ and $V=0.5$. 
Note that $N(\omega=\mu)$ on the Fermi surface 
is finite for any 
value of $t_y$, i.e., the system is metallic and the CO does not occur. 
It is surprising that such a small $V$ largely affects the SC
interaction vertex. 

In this section we have investigated the effect of charge fluctuation induced
by the nearest-neighbor repulsion $V$ on the SC symmetry. Our results indicate that
the symmetry of electron pair changes from $d_{x^2-y^2}$ to $d_{xy}$,
even if there still exists strong spin fluctuation and $V$ is as weak as
not accompanying CO. It suggests that the effect of $V$ on SC is not
negligible, even in the metallic system with strong spin fluctuation. We should
incorporate not only the shape of the Fermi surface but also the effect
of $V(\bm{q})$ into discussion.


\section{Conclusions}

\begin{figure}[htbp]
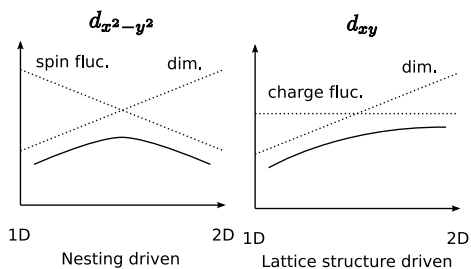

 \begin{center}
  \includegraphics[width=3cm]{fig14a.epsi} 
  \includegraphics[width=3cm]{fig14b.epsi} 
 \end{center}
 \caption{Schematic representation of the effect of dimensionality on SC
 fluctuation in Q1D $1/4$-filled system. 
 }
 \label{SpinCharge}
\end{figure}

In this paper we have investigated the superconductivity in Q1D system from
the non-perturbative point of view, focusing on the effect of
dimensionality on SC fluctuation. 
First we introduced the result of
the Hubbard model, which includes only the on-site repulsion $U$. 
In this case, it is found that 
the $d_{x^2-y^2}$-wave pair-field susceptibility, $\chi_{d{\rm SC}}$, 
is increased by $t_y$ in the small $t_y$ region, while the spin
fluctuation $\chi_{\rm s}(\bm{Q})$ is monotonically suppressed by $t_y$.
This behavior comes from the fact that the increase of $t_y$ has two different
effects; the deformation of the Fermi surface and the enhancement of
pair-hopping of electrons. 
The $d_{x^2-y^2}$-wave SC is most enhanced at the region where these two
factors are well balanced, as shown in the left panel of
Fig. \ref{SpinCharge} schematically. 
This behavior suggests that
the monotonic decline of $T_{\rm c}$ may not be the case. 
On the other hand, the pairing mechanism itself is the well-known spin
fluctuation assumed by many previous studies. It is
confirmed by the fact that the shape of the Fermi surface determines
whether the particle-particle interaction vertex $\Gamma$ for
$d_{x^2-y^2}$-wave pairing is attractive or not.

Next we considered the effect of nearest-neighbor repulsion $V$, 
focusing on the effect of charge fluctuation induced by $V$ on the
symmetry of electron pair. The expected behavior for $d_{xy}$-wave
pairing is shown in the right panel of Fig. \ref{SpinCharge}. 
We speculate 
that the charge fluctuation induced by $V$ is not
affected by the nesting property of the Fermi surface but the
geometrical lattice structure. 
In this case, the increase of $t_y$ simply makes a positive contribution
to SC because of 
the enhancement of the electron pair-hopping.
Our results support this picture, and in addition, indicate 
that 
$\Gamma$ 
for $d_{x^2-y^2}$-wave pairing is repulsive when $V$ is finite, even at
the region where strong spin fluctuation still exists and $V$ is as
small as not accompanying CO.
On the other hand, 
$d_{xy}$-wave
pairing is not so largely affected by $V$ and $\Gamma$ for $d_{xy}$-wave
pairing remains attractive. 
This indicates that the nesting property is not necessarily the crucial
factor for determining the symmetry of electron pair, even if the system
has a good nesting property. 
It may sound peculiar, especially from the perturbative point of view 
which focuses on the pairing mechanism mediated by the fluctuation of
the ordered phase next to the SC. However, 
the possibility of strong
cancellation between spin and charge fluctuations in $\Gamma$ would not be so
unrealistic. It suggests that we should incorporate the effect of long range Coulomb
interaction $V$ into the discussion of SC in order to determine the
symmetry of electron pair. 

Let us compare our results with experiment. 
We showed that the $\chi_{d{\rm SC}}$ is enhanced by increasing
$t_y$, whichever the symmetry of electron pair is $d_{x^2-y^2}$ or $d_{xy}$. 
It suggests that the SC phase in TM$_2$X salts have the possibility of
becoming wider, such as what is confirmed in (TMTTF)$_2$SbF$_6$
\cite{JPSJ.77.023701}. 
Furthermore, we also found that the $d_{xy}$-wave pairing becomes more
realistic than
$d_{x^2-y^2}$-wave pairing under the presence of nearest-neighbor repulsion
$V$. It indicates that the $d_{x^2-y^2}$-wave pairing which is assumed
in the 
previous studies of TM$_2$X-salts may need to be revised. In fact, not only the
TMTTF-salts, the TMTSF-salts showing the metallic behavior at ambient
pressure have the precursor of CO as the higher-energy peak of the
optical conductivity \cite{Science.281.1181}. This means that the
long-range Coulomb interaction is not negligible in not only TMTTF-salts 
but also TMTSF-salts.

Finally, let us comment on the temperature range of the present
calculation. We cannot study very 
low temperature region due to the severe sign problem. 
This could make trouble with the analysis of 
interaction vertex $\Gamma$; it is possible that 
the strengths of the spin and charge fluctuations which contribute to 
$\Gamma$ changes as temperature is lowered, 
since the spin susceptibility $\chi_{\rm s}(\bm{Q})$ grows rapidly 
while the charge susceptibility $\chi_{\rm c}(\pi,\pi)$ shows more
moderate behavior in the temperature range we calculated (not shown). 
Thus, 
the effect of $V$ on SC symmetry at lower temperatures still remains
open. We hope that more 
studies would be carried out both from theoretical and experimental
point of view. 

\section*{Acknowledgment}
We are grateful to T. Kato and Y. Uwatoko for helpful discussions.
The present work was financially supported by Grants-in-Aid for
Scientific Research on Priority Areas of Molecular Conductors
(No. 15073210) from the Ministry of Education, Culture, Sports, Science
and Technology, Japan (MEXT), and Next Generation Supercomputing
Project, Nanoscience Program, MEXT. Y.F. is supported by JSPS Research
Fellowships for Young Scientists. 
The computation in this work has been done using the
facilities of the Supercomputer Center, Institute for Solid State
Physics, University of Tokyo.


\begin{thebibliography}{99}
\bibitem{footnote01}
	 It should be noted that the charge fluctuation is also affected
	 by the shape of the Fermi surface. For example, 
	 $\chi_{\rm c}(\bm{q})$ is given as 
	 $\chi_{\rm c}(\bm{q})=\chi_{0}(\bm{q})/(1+V(\bm{q})\chi_0(\bm{q}))$
	 within RPA, which clearly shows that $\chi_{\rm c}(\bm{q})$ is
	 affected by the nesting property through
	 $\chi_0(\bm{q})$. In fact, out result of $\chi_{\rm c}(\bm{q})$
	 for $V=0$ slightly reflects the shape of the Fermi surface.
	 However, the effect of $V(\bm{q})$ is
	 stronger than that of $\chi_0(\bm{q})$. This leads to the
	 emergence of 
	 $\bm{q}=(\pi,\pi)$ charge disproportionation under the presence
	 of $V$ which reflects the lattice structure rather than the
	 nesting condition.
\end{thebibliography}
\end{document}